\documentclass[]{aastex631}

\usepackage{color}
\usepackage{hyperref}
\usepackage[normalem]{ulem}             

\renewcommand{\ion}[2]{#1\,{\sc #2}}

\shorttitle{Revised ionization rates}
\shortauthors{K. P. Dere}

\graphicspath{{./}{*}}

\begin{document}

\title{CHIANTI -- an atomic database for emission lines - Paper XVII:  Version 10.1, revised ionization and recombination rates and other updates}

\footnote{Released on July 17, 2023}

\author[0000-0003-1628-6261]{Kenneth P. Dere }
\affiliation{Department of Physics and Astronomy, George Mason University, 4400 University Drive, Fairfax, VA 22030, USA}

\author[0000-0002-4125-0204]{G. Del Zanna}
\affiliation{DAMTP, Center for Mathematical Sciences, University of Cambridge, Wilberforce Road, Cambridge, CB3 0WA, UK}

\author[0000-0001-9034-2925]{P. R. Young}
\affiliation{NASA Goddard Space Flight Center, Code 671,
  Greenbelt, MD 20771, USA}
\affiliation{Northumbria University, Newcastle Upon Tyne NE1 8ST, UK}

\author[0000-0002-9325-9884]{E. Landi}
\affiliation{Department of Climate, Space Sciences and Engineering, University of Michigan, Ann Arbor, MI, 48109}

\begin{abstract}

The CHIANTI atomic database provides sets of assessed data used for simulating spectral observations of astrophysical plasmas. This article describes updates that will be released as version~10.1 of the database. A key component of CHIANTI is the provision of ionization and recombination rates that are used to compute the ionization balance of a plasma over a range of temperatures.
Parameters for calculating the ionization rates of all stages of ions from H through Zn were compiled and inserted into the CHIANTI database in 2009.  These were based on all measurements that were available at the time and supplemented with distorted wave calculations.  Since then, there have been a number of new laboratory measurements for ions that produce spectral lines that are commonly observed.  Parameters have been fit to these new measurements to provide improved ability to reproduce the ionization cross sections and rate coefficients, and these are added to the database. CHIANTI 10.1 also includes new recombination rates for the phosphorus isoelectronic sequence, and 
the updated ionization and recombination rates have been used to calculate a new ionization equilibrium file. In addition, CHIANTI 10.1 has new electron collision and radiative datasets  for eight ions in the nitrogen and oxygen isoelectronic sequences, and updated energy level and wavelength data for six other ions.

\end{abstract}

\keywords{atomic data --- atomic processes --- Sun: UV radiation --- Sun: X-rays, gamma rays --- Ultraviolet: general --- X-rays:  general}

\section{Introduction} \label{sec:intro}

\begin{figure}[ht!]
\plotone{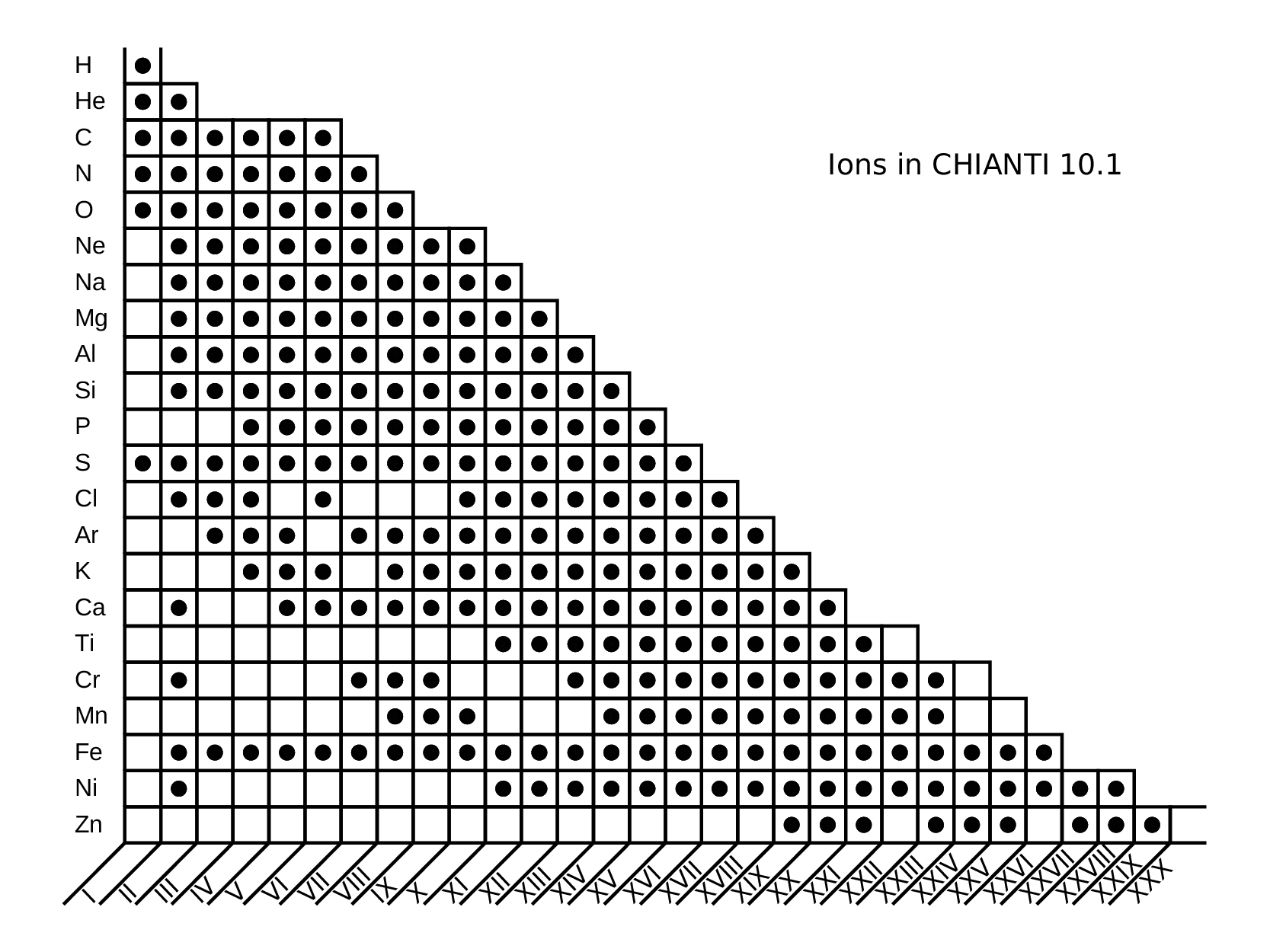}
\caption{A table showing the ions that are included in the CHIANTI version 10.1 database.}
\label{fig:chianti_ion_table}
\end{figure}

CHIANTI is an atomic database and software package used for modeling optically-thin emission from astrophysical plasmas. It has been an open-source project since the first release in 1996 \citep{paper1}, and the data and software are available at \url{https://chiantidatabase.org}. The most recent previous release is CHIANTI 10 \citep{paper16}, and the present article describes the updates and new features of the 10.1 release. \citet{2016JPhB...49g4009Y} and \citet{2020Atoms...8...46D} summarize the contents of the modern version of the database and provide applications.  The CHIANTI database contains a minimal set of information for all atoms and ions of all elements between H and Zn that includes descriptions of the ionization and recombination rates.  There are currently 283 ions in the database that contain more detailed information.  For these ions, there must be a file describing the energy levels, a file containing the wavelengths, weighted-oscillator strengths and the Einstein A coefficients (A-values) for the set of emitted lines, and file containing the information to reconstruct the electron collision strengths between ions.  These three files all refer to the same sets of energy levels.  There can be additional files describing autionization, level resolved ionization and recombination, and proton rates.  A table showing the current list of ions with detailed information is shown in Fig.~\ref{fig:chianti_ion_table}.

The spectral emission from ionized astrophysical plasmas is a key to our understanding of the processes that form and govern these regions.  A key to the analysis of these spectra is a knowledge of the ionization state.  For plasmas in a collisional ionization equilibrium, such as the solar corona, the equilibrium is largely controlled by electron collisional ionization and electron collisional recombination.  Collisional ionization consists of both direct ionization (DI) and excitation followed by autoionization (excitation-autoionization or EA).  Recombination also consists of two different processes, radiative recombination (RR) and dielectronic recombination (DR).  \citet{ioniz} analyzed a large number of measured ionization cross sections in order to make ionization cross sections  and their associated ionization rate coefficients readily available through the CHIANTI atomic database \citep{paper1, paper9, paper16}.  Since that analysis, a number of important ionization cross sections have been measured and published.  Here, these new measurements are analyzed in order to improve the calculation of ionization cross sections and rate coefficients from the data files in CHIANTI.

Recombination has been the focus of a project described by \citet{2003A&A...406.1151B} to systematically generate new DR and RR data for entire isoelectronic sequences. Previous versions of CHIANTI have used these rates for sequences from hydrogen to silicon. New calculations for the phosphorus sequence were presented by \citet{bleda_pseq} and these are discussed in Sect.~\ref{sec:pseq}. A new ionization balance is calculated from the updated ionization and recombination rates and is discussed in Sect.~\ref{sec:ioneq}.

The level balance within an ion is determined by electron collision strengths, radiative decay rates and energy levels, and updates to these datasets are described in Sect.~\ref{sec:models}.

\section{Revised ionization and recombination rate coefficients} \label{sec:ionx}

\subsection{Approach} \label{sec:approach}

The approach used by \citet{ioniz} was to compile a set of parameters for each ion to be able to reproduce the available ionization cross sections measurements.  The first priority was to use measured cross sections and when these were not available, cross sections were calculated with the Flexible Atomic Code  \citep[FAC:][]{fac}.  In this paper, new measurements for 13 ions are examined.  In all cases, this leads to new fits to laboratory measurements or adjustments to existing theoretical cross sections.

\citet{ioniz} developed a scaling for ionization cross sections similar to that of \citet{burly} for collision strengths.  We define a scaled energy $U$, 
\begin{equation}
U = 1 - \frac{\ln f}{\ln(u -1 +f)}
\label{equ:btE}
\end{equation}
and a scaled cross section $\Sigma$,
\begin{equation}
\Sigma = \frac{u \, \sigma \, I^2}{\ln u + 1}
\label{equ:btX}
\end{equation}
where $u = E/I$, $E$ is the energy of the incident electron, $I$ is the ionization potential, $\sigma$ is the ionization cross-section, and $f$ is a scaling parameter that is selected by the data assessor.

The scaled energy $U$ varies from 0 at  $I$ to unity for infinite energy.  The scaling parameter adjusts the placement of the scaled energies between these two limits.  The scaling is chosen to spread the region just above the ionization potential so that the variation of $\Sigma$ becomes more apparent and the fit to the measured values easier to evaluate.  We refer to this process as BTI (Burgess-Tully-Ionization) scaling.
A spline fit to the $\Sigma$ values is performed after $f$ has been chosen, and the spline values are stored in the {\tt diparams} CHIANTI file for each ion.  The information in these files, such {\tt fe\_24.diparams}, allows the direct ionization rate of \ion{Fe}{xxiv} to be calculated as a function of temperature.  Typically nine or more spline nodes are used.  Since the low energy behavior of $\Sigma$ is very apparent, it is possible to determine if additional spline nodes are needed.  This is often the case at the energy region just above the ionization potential, that is important for determining the electron ionization rate coefficient.  This is illustrated in Sect.~\ref{sec:rate}.

\begin{figure}[ht!]
\plotone{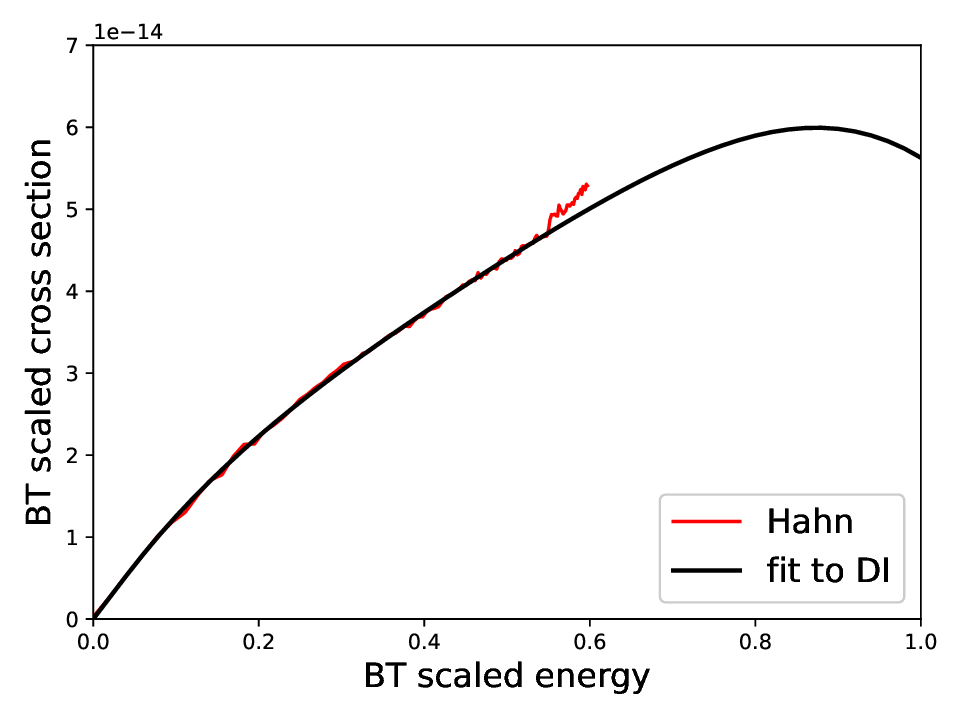}
\caption{BTI scaled ionization cross sections of \ion{S}{xiii}.  The label {\it Hahn} refers to \citet{hahn_s_13}, and {\it present} to the results reported here.}
\label{fig:s_13_btcross}
\end{figure}

An example of fitting BTI scaled measured cross sections is shown in Fig.~\ref{fig:s_13_btcross} for \ion{S}{xiii}.  Here the BTI scaled measurements of \citet{hahn_s_13} are shown together with the current fit to the BTI scaled measurements below 2360 eV ($U$ = 0.55).  The fit to the measurements provides the parameters to reproduce the direct ionization cross sections of the ion in question.  The high energy limit, $U$ = 1, is given by the Bethe cross section as discussed by \citet{ioniz}.  
The deviation of the measurements from the fit between $U=0.55$ and $U=0.60$ shown in Fig.~\ref{fig:s_13_btcross} is due to the excitation-autoionization (EA) component. \citet{ioniz} computed the EA parameters using FAC and in the present work adjustments are made to the threshold and/or the amplitude to best match the new measurements.  As will be described in $\S$ \ref{sec:s_13} the parameters for the EA component calculated with CHIANTI have been modified to provide a good reproduction of the experimental cross section at these energies.

The procedure for each ion examined is discussed in more detail later in the following Sections. 

Finally, we note that \cite{dufresne_delzanna:2019} and \cite{dufresne_etal:2020} recently calculated ionization cross sections for all ions of carbon and oxygen using FAC and AUTOSTRUCTURE \citep{badnell:2011}, and provided comparisons to available experimental data. Cross sections for collisional ionization from ground states were in most cases close (within 15\%) to the Dere (2007) ones.
The calculations were focused on ionization from metastable levels, and will be made available in a future CHIANTI release.

\subsection{The beryllium isoelectronic sequence}\label{sec:beseq}

\subsubsection{O V} \label{sec:o_5}

Crossed beams measurements of the ionization cross sections of \ion{C}{iii}, \ion{N}{iv}, and \ion{O}{v} made at the Oak Ridge National Laboratory have been presented by \citet{fogle_beseq}.  Prior measurements of these ions have been reported by \citet{falk1983}.  They were considered in the analysis of \citet{ioniz} but these cross sections showed considerable amplitudes at energies below the ground state ionization potential, indicating a large fraction of ions in metastable levels.

The measurements of \citet{fogle_beseq} for \ion{C}{iii} and \ion{N}{iv} show considerable cross sections below the ionization potential, indicating a beam containing a significant population in the 2s2p $^3$P$_{0,1,2}$ metastable levels.  They estimate that the fraction of ions in the metastable levels is 0.46, 0.30 and 0.24 for \ion{C}{iii}, \ion{N}{iv}, and \ion{O}{v}, respectively. The measurements of \citet{fogle_beseq} for \ion{C}{iii} and \ion{N}{iv} show significant values of the cross section below the ground state ionization potential and these ions  will not be included in this analysis.  The \ion{O}{v} measurements of \citet{fogle_beseq} do not indicate the presence of significant populations of excited levels.

\begin{figure}[ht!]
\plotone{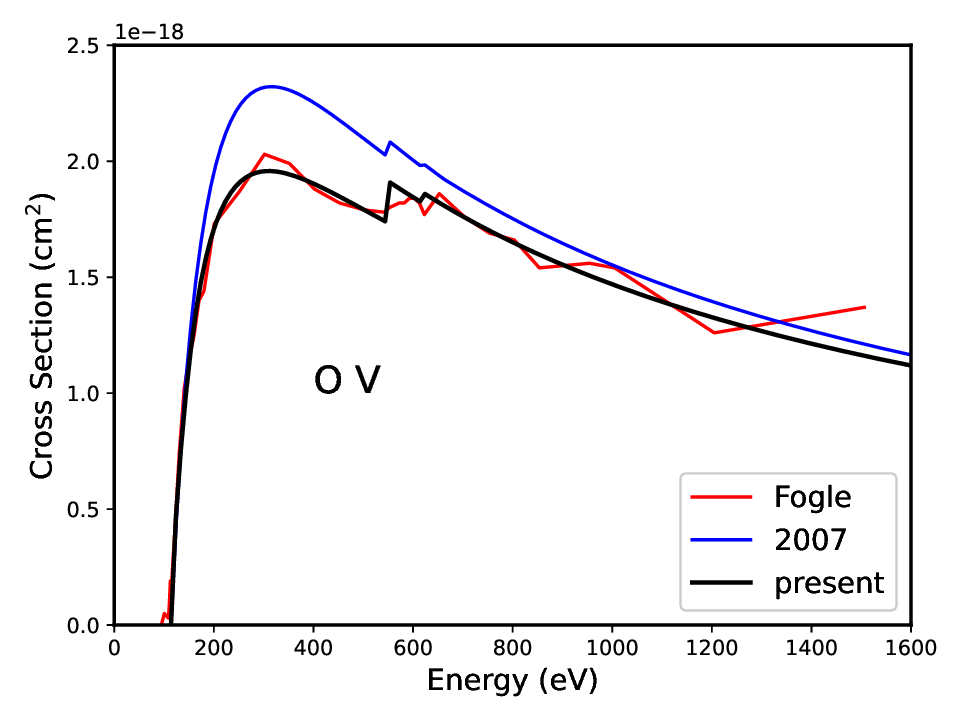}
\caption{Ionization cross sections of \ion{O}{v}.  The label {\it Fogle} refers to \citet{fogle_beseq}, {\it 2007} refers to \citet{ioniz}, and {\it present} to the results reported here.}
\label{fig:o_5_cross}
\end{figure}

As in $\S$ \ref{sec:s_13}, the direct ionization cross sections are represented by a fit to the BTI scaled cross sections of \citet{fogle_beseq} below about 520 eV where EA processes become significant.    The EA cross sections of \citet{ioniz} have been increased by a factor of 2.5 to match the measurements but the two EA thresholds remain at 552 and 620 eV.  The \ion{O}{v} measurements of \citet{fogle_beseq}, \citet{ioniz} and the current analysis are presented in Fig.~\ref{fig:o_5_cross}.  The peak cross sections near about 310 eV show that the current values are reduced by a factor of about 0.84 from the FAC values of \citet{ioniz}.

We note that earlier experimental data had significant limitations. 
The  \citet{crandall1979} and \citet{falk1983} experimental results were in agreement with each other, but they were not able to determine the metastable fraction present. \citet{loch2003} produced similar experimental values, although significant readings were present below threshold, indicating a linear offset in readings or contamination in the beam. All of the above cross sections were higher than theory by 30-50\%. The \citet{fogle_beseq} experimental cross sections  were in good agreement with their R-Matrix calculations at low and high energies, while at the peak they were about 25\% below. The \citet{dufresne_etal:2020} calculations are in good agreement with the \citet{fogle_beseq} ones.

The adjustments to the EA cross sections are made by a visual comparison between the experimental results and the FAC calculations.  These can consist of adjustments to both the magnitude and the threshold by editing the tabulated directly by hand.  No new FAC calculations were performed.  The same method has been applied to the EA adjustments later in the paper for a number of ions.  Considering the errors in the measurements and the simplicity of the model calculations, this procedure is appropriate.

\subsubsection{S XIII} \label{sec:s_13}

Storage ring measurements of \ion{S}{xiii} ionization cross sections have been presented by \citet{hahn_s_13}.  A major problem with experimental measurements of ions in the beryllium isoelectronic sequence is that the ion beams can contain a significant population of ions in the 2s2p $^3$P$_0$ metastable level.  This results in cross section measurements that show an enhanced cross section below the ionization potential of the ground level.  As the authors explain, they used a beam of the $^{33}$S isotope that has a nuclear spin.  ``The resulting hyperfine interaction induces a mixing of the $^3$P$_1$ and $^3$P$_0$ levels, decreasing the lifetime of the $^3$P$_0$ level.''  The beam of \ion{S}{xiii} ions were kept in the ion storage ring long enough that the metastable levels decay to the ground state.

\begin{figure}[ht!]
\plotone{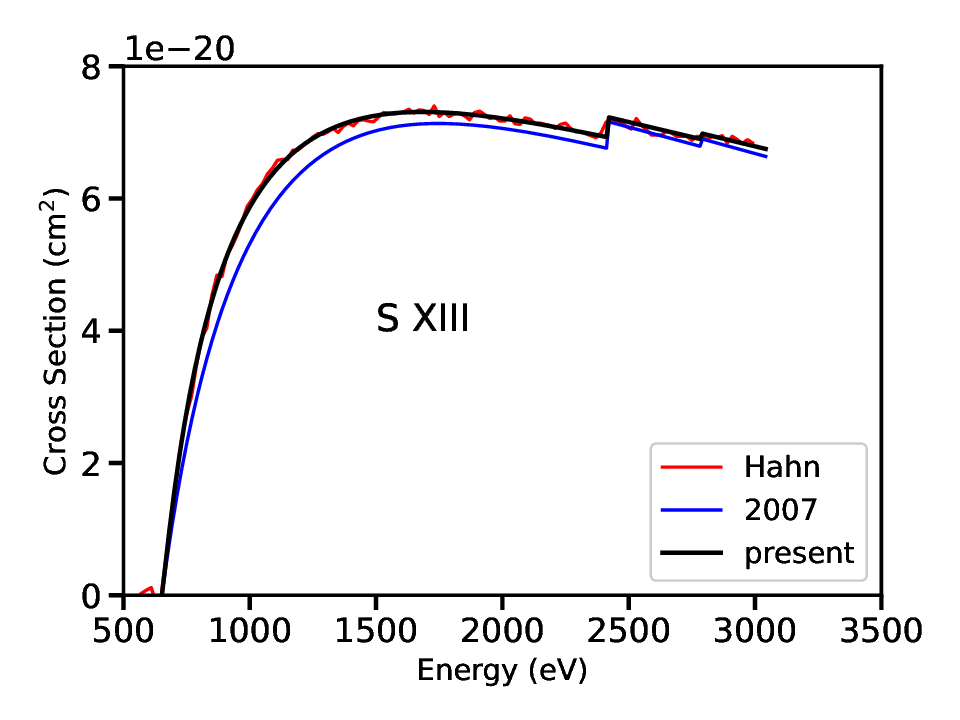}
\caption{Ionization cross sections of \ion{S}{xiii}.  The label {\it Hahn} refers to \citet{hahn_s_13}, {\it 2007} refers to \citet{ioniz}, and {\it present} to the results reported here.}
\label{fig:s_13_cross}
\end{figure}

Fig. \ref{fig:s_13_cross} shows the present and previous cross sections together with the values of \citet{hahn_s_13}.  The original FAC calculations indicated excitation-autoionization contributions at energies of 2412 and 2786 eV for excitations to the 1s$\,$2s$^2$$\,$2p and 1s$\,$2s$^2$$\,$3{\it l} levels, respectively.  The energies of the EA contributions are consistent with the measurements but the magnitudes of both have been decreased by a factor of 0.75 to be consistent with the measurements.  The ratio of the maximum cross section at about 1600 eV is about 1.03 for the present results to the 2007 FAC values and the ratio at a lower energy of 1000 eV is about 1.10.

\subsection{The boron isoelectronic sequence:  Mg VIII} \label{sec:mg_8}

\citet{hahn_mg_8} reports measurements of the ionization cross sections of \ion{Mg}{viii} performed at the heavy ion storage ring TSR at the Max-Planck-Institut f\"{u}r Kernphysik. They compare their measurements of the FAC cross sections to \citet{ioniz} and find that, near the peak of the cross section, their measurements are below the FAC calculations but within the error bars.  At energies just above threshold near 300~eV, their measurements are above the FAC values and at the very lowest energies the differences exceed their error bars.  While the earlier FAC calculations were not too far from the new measurements, it is worthwhile to perform a fit to the \citet{hahn_mg_8} measurements.  The FAC calculations of \citet{ioniz} only considered direct ionization from the 2s and 2p levels and no EA contributions.

\begin{figure}[ht!]
\plotone{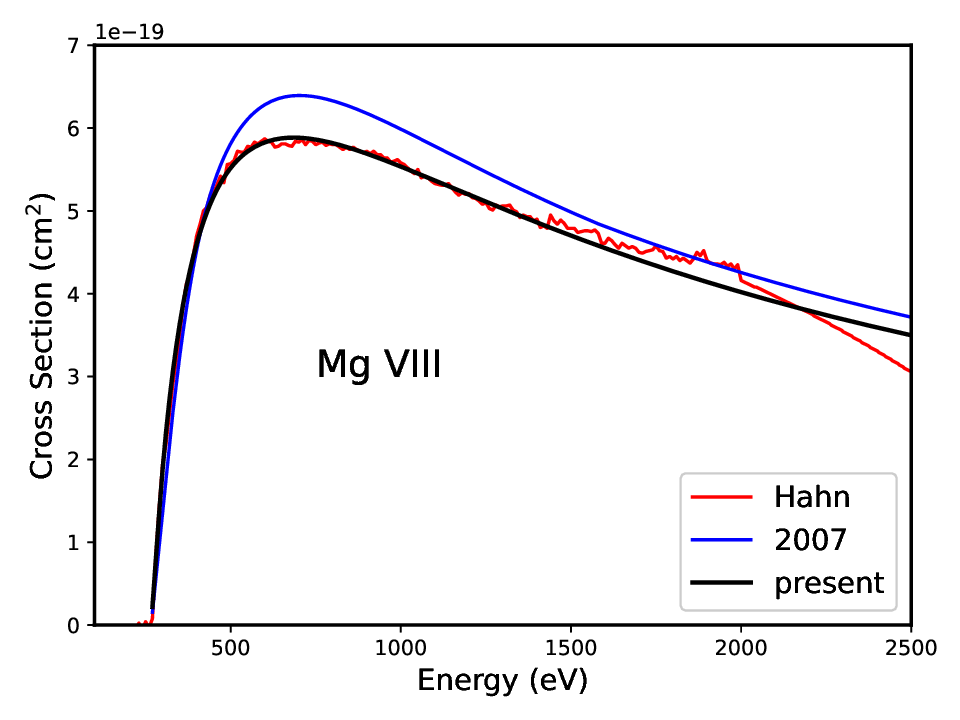}
\caption{Ionization cross sections of \ion{Mg}{viii}.  The label {\it Hahn} refers to \citet{hahn_mg_8}, {\it 2007} refers to \citet{ioniz}, and {\it present} to the results reported here.} 
\label{fig:mg_8_cross}
\end{figure}

The BTI scaled experimental cross sections have been fit in the manner described in $\S$ \ref{sec:s_13} with particular attention paid to the measurements just above threshold.  The \ion{Mg}{viii} measurements of \citet{hahn_mg_8}, the FAC cross sections of \citet{ioniz} and the current fits to the experimental data are presented in Fig. \ref{fig:mg_8_cross}.  As the authors suggest, ``the apparent structure in the experimental results near 600, 1400 and 1600 eV are due to run to run variations in the background levels."  At peak cross sections near about 700 eV the current values are reduced by a factor of about 0.92 from the FAC values of \citet{ioniz}.  However, at an energy of about 300 eV the current values are about a factor of 1.3 above the FAC cross sections.

\subsection{The fluorine isoelectronic sequence:  Fe XVIII} \label{sec:fe_18}

The ion \ion{Fe}{xviii} provides a number of strong lines between 14 and 18~\AA.  These make important contributions to observations by Chandra and other missions that observe at X-ray wavelengths.  \citet{hahn_fe_14_17_18}  reported measurements of the cross sections of \ion{Fe}{xiv}, \ion{Fe}{xvii} and \ion{Fe}{xviii} at the heavy ion storage ring TSR, and  \ion{Fe}{xviii} is discussed here.

\begin{figure}[ht!]
\plotone{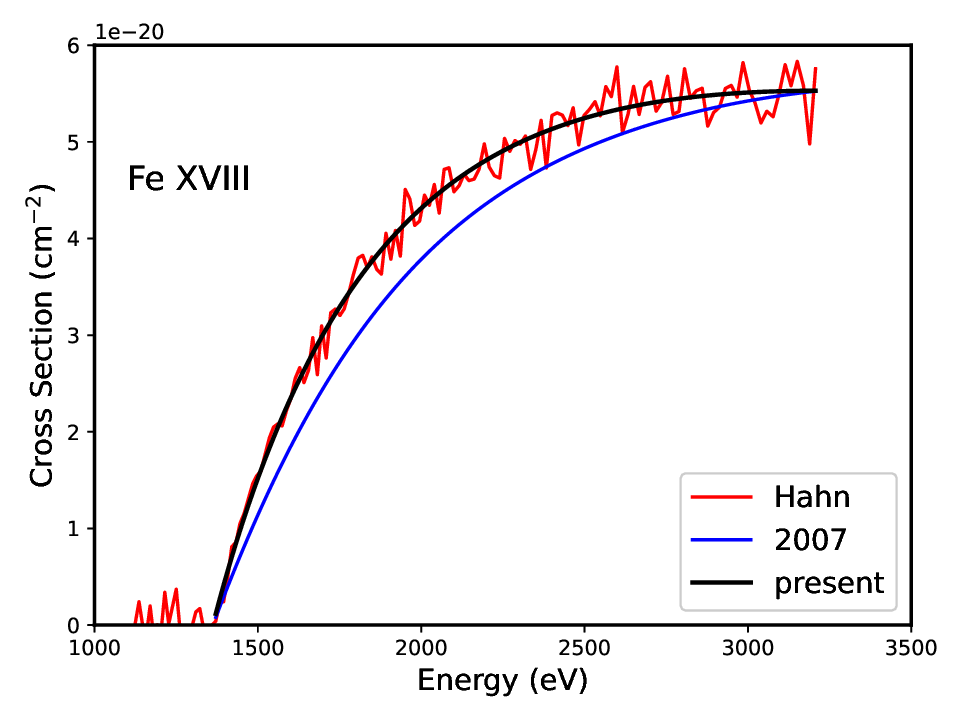}
\caption{Ionization cross sections of \ion{Fe}{xviii}.  The label {\it Hahn} refers to \citet{hahn_fe_14_17_18}, {\it 2007} refers to \citet{ioniz}, and {\it present} to the results reported here.} 
\label{fig:fe_18_cross}
\end{figure}

The BTI scaled experimental cross sections of \citet{hahn_fe_14_17_18} have been fit in the manner described in $\S$ \ref{sec:approach} with particular attention paid to the measurements just above threshold.  A comparison of the measurements of \citet{hahn_fe_14_17_18}, \citet{ioniz} and the present results are shown in Fig. \ref{fig:fe_18_cross}.  Neither do the measurements of the ionization cross sections of \ion{Si}{vi}, also in the fluorine sequence, by \citet{thompson_si5_si6} indicate any significant EA contributions.  The FAC calculations of \citet{ioniz} only considered direct ionization from the 2s and 2p levels and no EA contributions.

At energies of about 2000 eV, somewhat above threshold, the present cross section is about a factor of 1.13 greater than the 2007 values, but the ratio becomes unity at 3200 eV and 0.94 at about 5000 eV.

\subsection{The neon isoelectronic sequence:  Fe XVII} \label{sec:fe_17}

\ion{Fe}{xvii} is a strong contributor to the X-ray spectrum in the wavelength range 10 to 18 \AA\ and at extreme ultraviolet (EUV) wavelengths between 200 and 300 \AA.  A number of the EUV lines have been observed in solar spectra between 180 and 400 \AA\ \citep{warren_eis_hiT, delzanna_fe_17}.

\begin{figure}[ht!]
\plotone{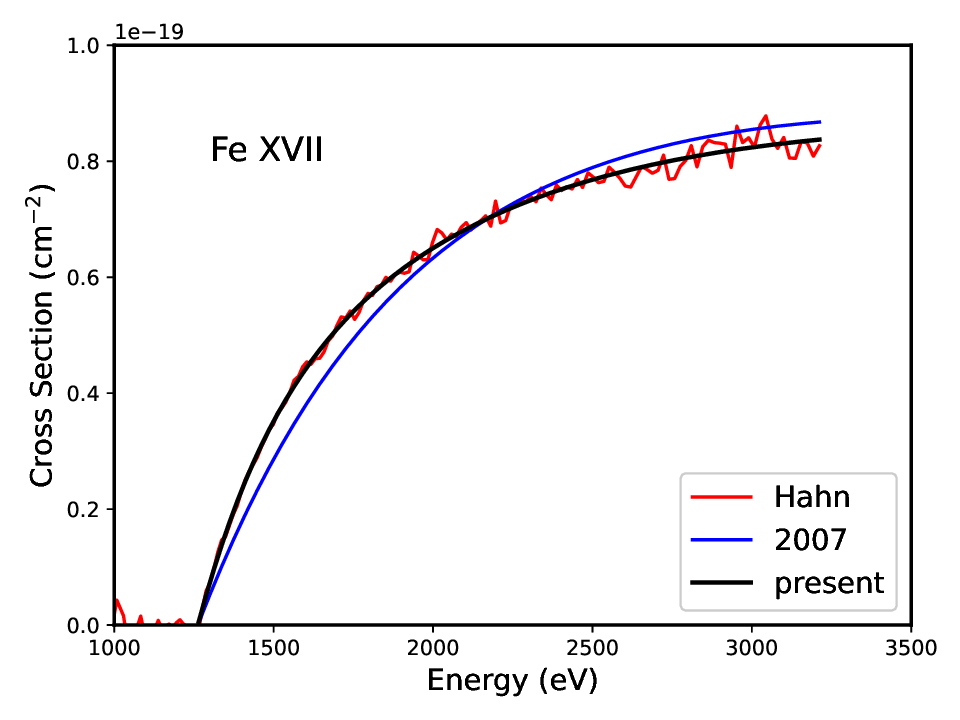}
\caption{Ionization cross sections of \ion{Fe}{xvii}.  The label {\it Hahn} refers to \citet{hahn_fe_14_17_18}, {\it 2007} refers to \citet{ioniz}, and {\it present} to the results reported here.}
\label{fig:fe_17_cross}
\end{figure}

Measurements of the ionization cross section of \ion{Fe}{xvii} are described by \citet{hahn_fe_14_17_18}.  A comparison of these measurements for \ion{Fe}{xvii}, those of \citet{ioniz} and the present results are shown in Fig. \ref{fig:fe_17_cross}.  The FAC calculations of \citet{ioniz} only took into account direct ionization from the 2p shell.  The new fits to the measured cross sections are able to reproduce the measurements quite well.

\subsection{The magnesium isoelectronic sequence:  Fe XV}  \label{sec:fe_15}

\begin{figure}[ht!]
\plotone{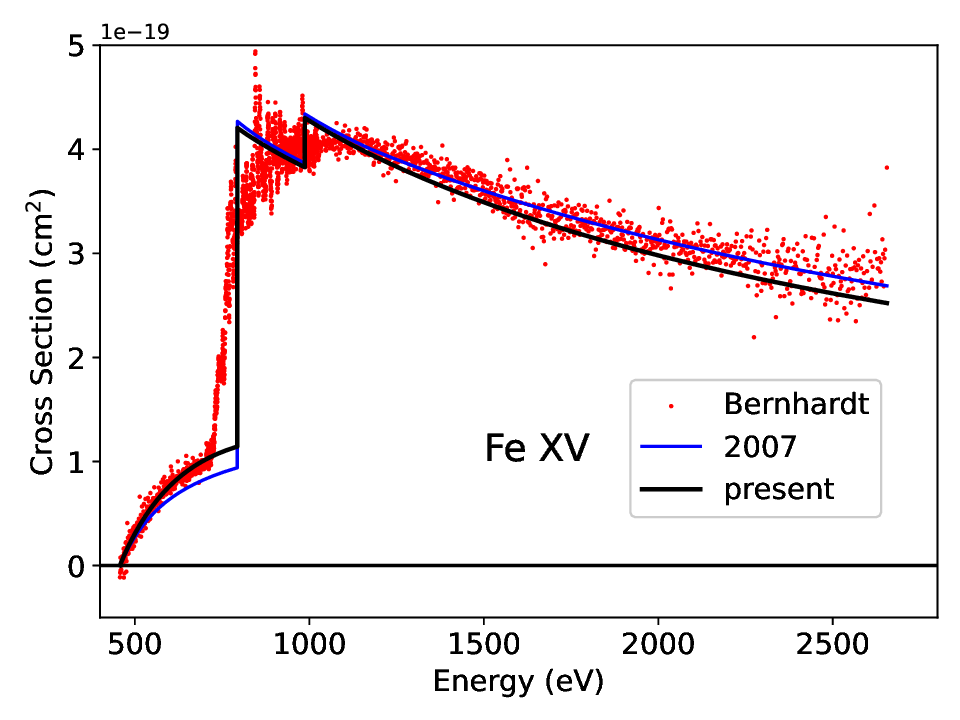}
\caption{Ionization cross sections of \ion{Fe}{xv}.  The label {\it Bernhardt} refers to \citet{bernhardt_fe_15}, {\it 2007} refers to \citet{ioniz}, and {\it present} to the results reported here.}
\label{fig:fe_15_cross}
\end{figure}

\ion{Fe}{xv} produces a strong line at 284 \AA\ in the EUV spectrum.  In addition, it also provides density-sensitive line pair ratios such as 234 \AA\ to 244 \AA\ \citep{cowan_widing}.  The ionization cross-section of \ion{Fe}{xv} consists of both DI and EA components and the \citet{ioniz} cross sections are based on calculations with FAC.  The EA components include transitions from the n=2 levels to the n=3 and n=4 levels and the various transitions are grouped into two EA components with excitation thresholds at 793 and 987 eV.  These compare to the ionization potential of 457 eV.  Laboratory measurements of the ionization cross sections of \ion{Fe}{xv} are reported by \citet{bernhardt_fe_15} and show both DI and EA components although the measured EA components appear more complex than the two component description.  In addition, \citet{bernhardt_fe_15} state that there is evidence of resonant processes that occur just before the major EA components.  These are not included in the present model described here.  The CHIANTI ionization cross sections have been revised by fitting the DI component to the \citet{bernhardt_fe_15} cross sections below 793 eV.  The EA component to the n=3 levels has been reduced by a factor of 0.92.  The measurements of \citet{bernhardt_fe_15} together with the \citet{ioniz} values and the present results are shown in Fig. \ref{fig:fe_15_cross}.  At energies just below the first EA threshold, the ratio of the present cross section to that of \citet{ioniz} is about 1.22.  At energies where the EA component is present, the ratio is about 0.99 at 1000 eV and 0.95 at 2000 eV.  \citet{bernhardt_fe_15} also provide parameters to reproduce their calculations of the ionization rate coefficient.  The rates of \citet{bernhardt_fe_15} are very close to those of the revised parameters.  The ratio of the present rates to those of \citet{bernhardt_fe_15} is 0.83 at 10$^5$ K, 0.91 at 2.2 $\times$ 10$^6$ K (the temperature of maximum ionization of  \ion{Fe}{xv}) and 0.93 at 3 $\times$ 10$^6$ K.

\subsection{The aluminum isoelectronic sequence:  Fe XIV}  \label{sec:fe_14}

\ion{Fe}{xiv} produces the strong ``green" line at 5304 \AA\ that is often observed with solar coronagraphs.  It is also responsible for a number of strong lines at EUV wavelengths.  Intensity ratios of some pairs of line are also sensitive indicators of electron densities between 10$^9$ and 10$^{11}$ cm$^{-3}$ \citep{dere_flare_densities}.

\begin{figure}[ht!]
\plotone{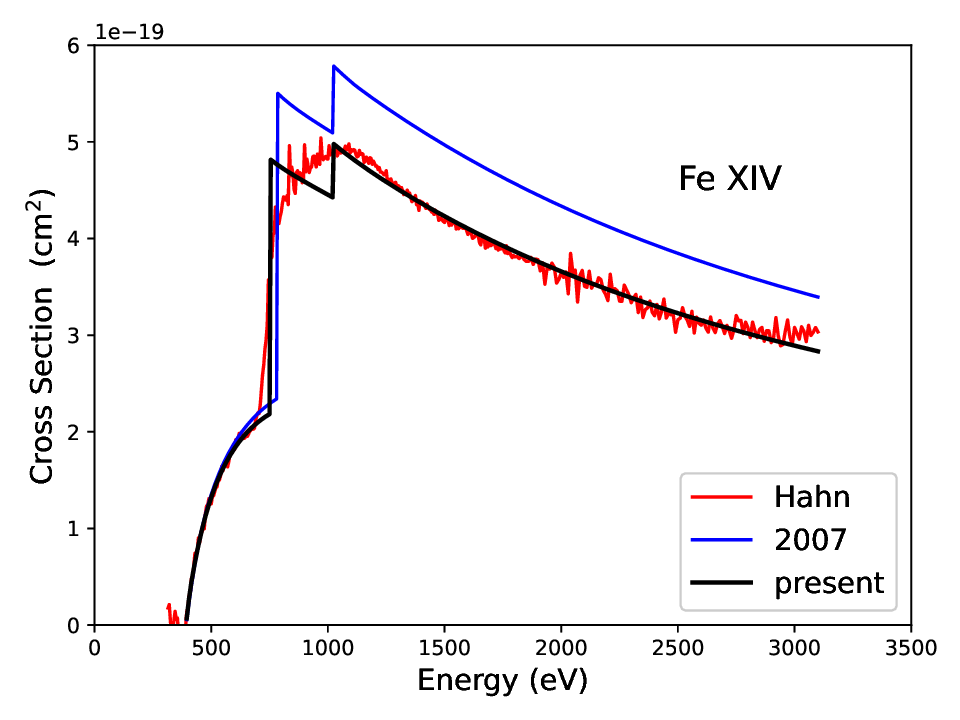}
\caption{Ionization cross sections of \ion{Fe}{xiv}.  The label {\it Hahn} refers to \citet{hahn_fe_14_17_18}, {\it 2007} refers to \citet{ioniz}, and {\it present} to the results reported here.}
\label{fig:fe_14_cross}
\end{figure}

As with \ion{Fe}{xv}, the \ion{Fe}{xiv} cross section includes both DI and EA components, and both are prominent in both the measurements of \citet{hahn_fe_14_17_18} and the FAC cross sections of \citet{ioniz}.  A comparison of the measurements of \citet{hahn_fe_14_17_18}, \citet{ioniz} and the present results are shown in Fig. \ref{fig:fe_14_cross}.  The present values consist of a fit to the DI measured cross sections below 783 eV where the EA components become apparent.  The FAC cross sections of \citet{ioniz} included excitations of a 2s electron to the n=3 and n=4 levels.  The new EA cross sections consist of adjustments to the EA threshold for the n=3 excitations and the magnitudes for both components are multiplied by a factor of 0.8.

\subsection{The silicon isoelectronic sequence:  Fe XIII}  \label{sec:fe_13}

\ion{Fe}{xiii} produces a number of strong spectral lines at EUV wavelengths.  Some of these line pairs are useful diagnostics of electron densities between 3 $\times$ 10$^8$ and 3 $\times$ 10$^{10}$ cm$^{-3}$ \citep{flower_nussbaumer_fe13}.

\begin{figure}[ht!]
\plotone{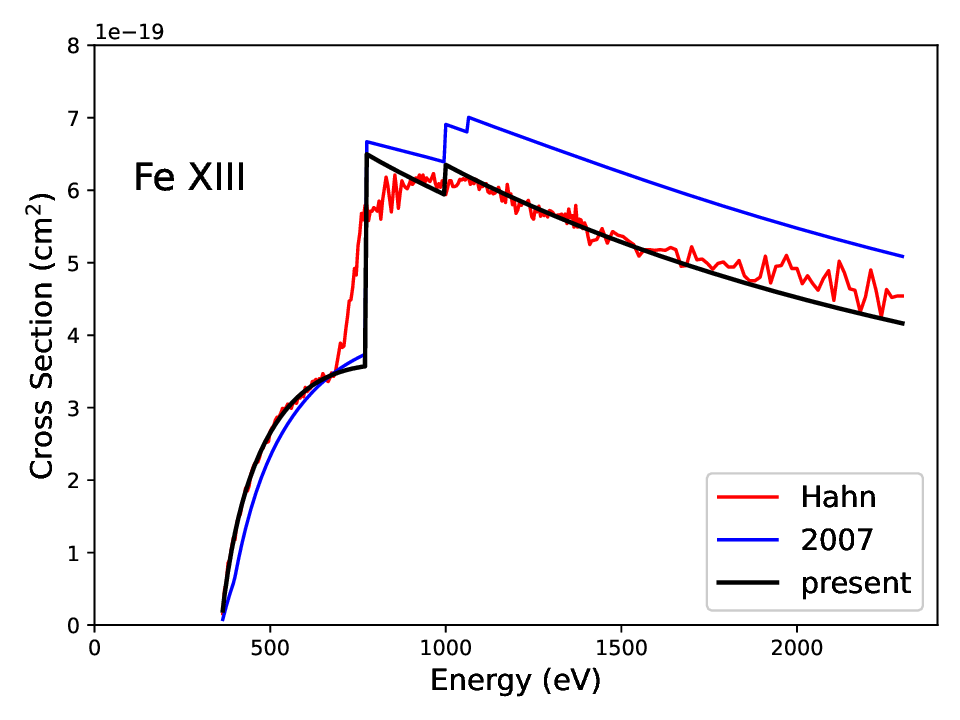}
\caption{Ionization cross sections of \ion{Fe}{xiii}.  The label {\it Hahn} refers to \citet{hahn_fe_13, hahn_fe_13_add}, {\it 2007} refers to \citet{ioniz}, and {\it present} to the results reported here.}
\label{fig:fe_13_cross}
\end{figure}

A comparison of the measurements of \citet{hahn_fe_13, hahn_fe_13_add}, \citet{ioniz} and the present results are shown in Fig. \ref{fig:fe_13_cross}.  As with \ion{Fe}{xiv}, the cross section consists of strong DI and EA components.  The DI fit parameters are arrived at by using the measurements below the first EA threshold at 775 eV.  The EA components of the FAC cross sections included excitations from the n=2 levels to the n = 3, 4, and 5 levels.  A better fit was obtained for the present cross section if the excitation to the n=5 levels was removed, the excitation to the n=4 levels reduced by a factor of 0.8 and the transition to the n=3 levels unchanged.  At 400 eV, just above the threshold, the ratio of the present cross section to that of the 2007  cross section is about 2.0, at 500 eV the ratio is about 1.17, at 600 eV the ratio is about 1.05, and at 700 eV the ratio is about 0.99.  Above 780 eV where the EA cross sections occur, the ratio of the present cross sections to the 2007  values is about 0.8.  These ratios are important for understanding the changes in the ionization rate coefficients discussed in $\S$ \ref{sec:rate}.

\subsection{The phosphorus isoelectronic sequence:  Fe XII} \label{sec:fe_12}

\ion{Fe}{xii} also produces a number of strong lines at EUV wavelengths.  Some are strong enough that they comprise the primary component of narrow band EUV images such as the Extreme-ultraviolet Imaging Telescope (EIT) on SOHO and the Atmospheric Imaging Assembly (AIA) on the Solar Dynamics Observatory (SDO).  A number of \ion{Fe}{xii} line pairs form ratios that are reliable diagnostics of electron densities between 10$^8$ and 10$^{11}$ cm$^{-3}$.   They have been observed with the  XUV Slitless Spectrograph on Skylab \citep{dere_flare_densities} and with the Extreme-ultraviolet Imaging Spectrometer (EIS) on the Hinode satellite \citep{dere_eis_qs, warren_qs}.  

\begin{figure}[ht!]
\plotone{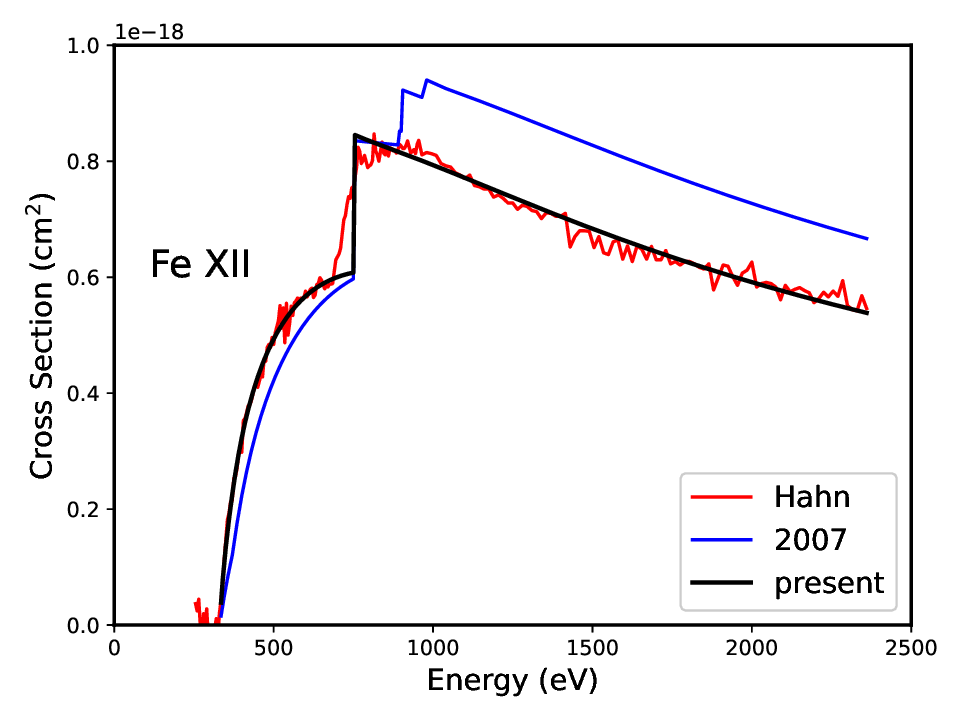}
\caption{Ionization cross sections of \ion{Fe}{xii}.  The label {\it Hahn} refers to \citet{hahn_fe_12}, {\it 2007} refers to \citet{ioniz}, and {\it present} to the results reported here.}
\label{fig:fe_12_cross}
\end{figure}

A comparison of the measurements of \citet{hahn_fe_12}, \citet{ioniz} and the present results are shown in Fig. \ref{fig:fe_12_cross}.  The DI FAC cross sections of \citet{ioniz} took into account ionization from the n=2 levels and the 3s and 3p levels.  These have been replaced with a fit to the \citet{hahn_fe_12} measurements below the first EA threshold at 753 eV.  The previous EA cross sections considered excitations from the n=2 levels to the n=3, 4, and 5 levels.  The comparison in Fig. \ref{fig:fe_12_cross} indicates that measurements are best matched if the excitations to the n=4 and 5 levels are removed.  Otherwise, it has not been necessary to change the energy of the n=3 threshold or magnitude.

Just above threshold the ratio of the present cross section to that of \citet{ioniz} is about 2.5; at 400 eV, the ratio is about 1.4; and at 600 eV, it is about 1.08.  Above 1000 eV where the EA cross sections occur, the ratio  is about 0.8.  Again, these ratios are important for understanding the changes in the ionization rate coefficients discussed in $\S$ \ref{sec:rate}.

\subsection{The sulfur isoelectronic sequence:  Fe XI} \label{sec:fe_11}

\begin{figure}[ht!]
\plotone{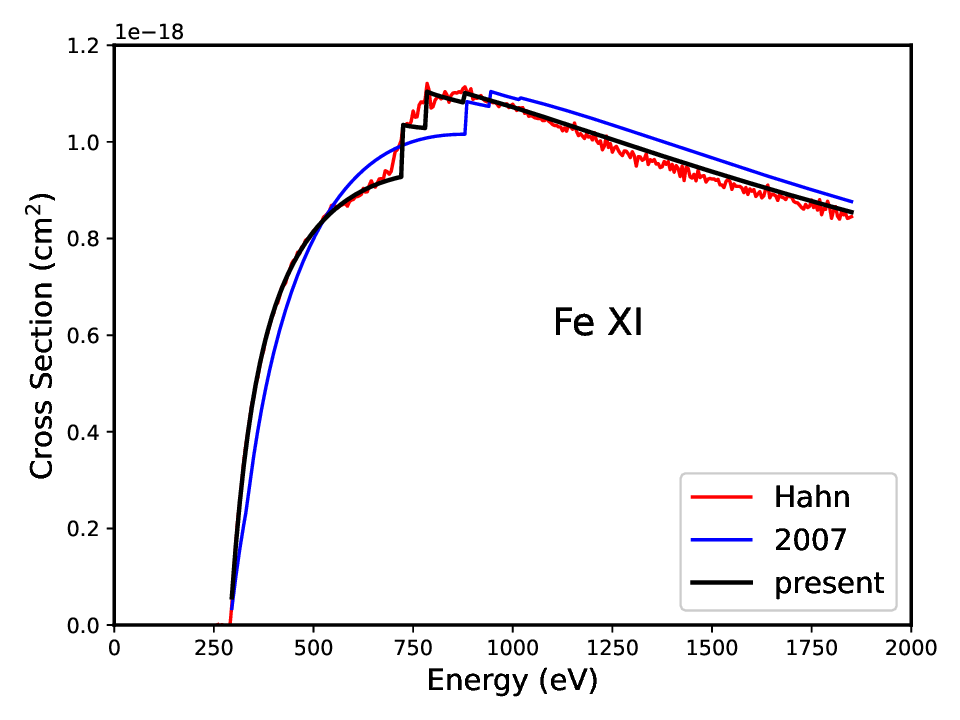}
\caption{Ionization cross sections of \ion{Fe}{xi}.  The label {\it Hahn} refers to \citet{hahn_fe_10_11}, {\it 2007} refers to \citet{ioniz}, and {\it 2022} to the results reported here.}
\label{fig:fe_11_cross}
\end{figure}

The ion \ion{Fe}{xi} contributes a number of strong lines to the EUV spectrum.  It also provides density sensitive line ratios that are useful in the density range between 10$^9$ and 10$^{11}$ cm$^{-3}$ \citep{dere_serts_qs_densities}.  A comparison of the measurements of \citet{hahn_fe_10_11}, \citet{ioniz} and the present results are shown in Fig. \ref{fig:fe_11_cross}.  The FAC EA cross sections included excitations from the n=2 to the n=4 and 5 levels.  The summed rates have been collected to form three EA cross sections.  The threshold energies have been adjusted to reproduce the measured cross sections.  For the 2007 EA cross sections, the lowest energy threshold was 882 eV and this has been moved to 721 eV.  The other two thresholds have been adjusted by a similar amount.  The magnitudes of the three original EA cross sections have been scaled by factors of 1.3, 4.0 and 2.0.  The ratio of the present cross sections to the 2007 values have been evaluated.  At 300 eV, just above the threshold, the ratio is about 1.6, at 500 eV the ratio is 1.01 at 700 eV, just before the EA cross sections occur, the ratio is about 0.96. At energies above 1000 eV, the ratio is about 0.97.

\subsection{The chlorine isoelectronic sequence:  Fe X} \label{sec:fe_10}

\ion{Fe}{x} produces a number of lines at EUV wavelengths as well as the coronal ``red" line at optical wavelengths.  Some EUV line pairs form ratios that are sensitive to electron densities between 3 $\times$10$^8$ and 10$^{11}$ cm$^{-3}$.

\begin{figure}[ht!]
\plotone{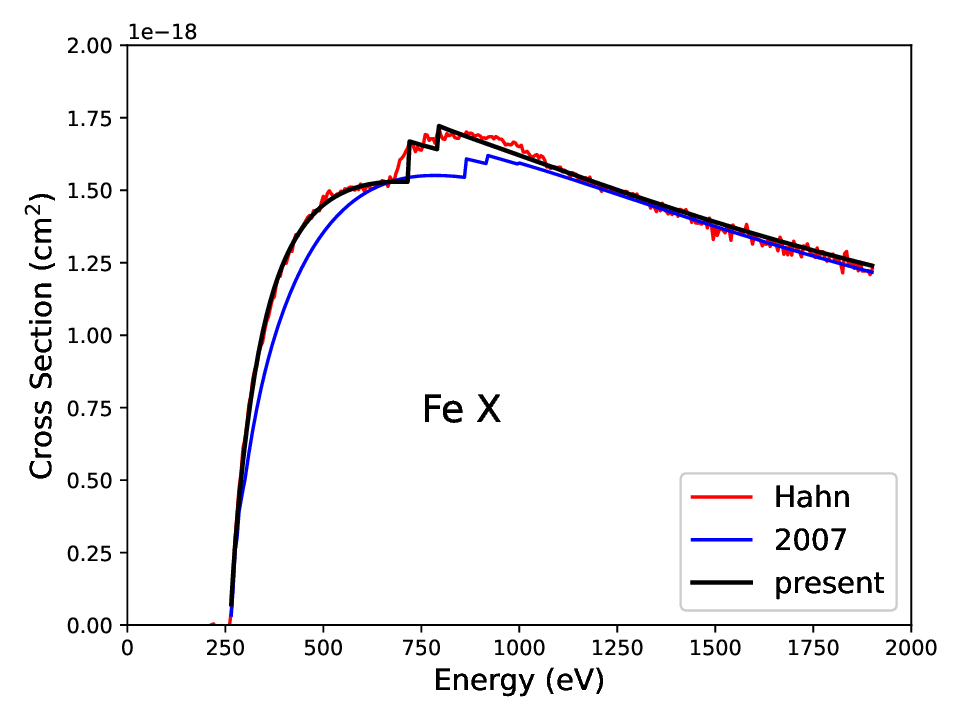}
\caption{Ionization cross sections of \ion{Fe}{x}.  The label {\it Hahn} refers to \citet{hahn_fe_10_11}, {\it 2007} refers to \citet{ioniz}, and {\it present} to the results reported here.}
\label{fig:fe_10_cross}
\end{figure}

The ionization cross sections of \ion{Fe}{x} have been measured by \citet{hahn_fe_10_11} and a comparison of these measurements, the FAC cross sections of \citet{ioniz} and the present results are shown in Fig. \ref{fig:fe_10_cross}.  From this figure, it is clear that both the DI and EA components of the original FAC calculations need to be modified to accurately reproduce the measured values.  The original FAC calculations took into account DI ionization from the n=2 and n=3 levels.  The EA components included excitation from the n=2 levels to the n=4 and n=5 levels and excitations from the n=3 to the n=4, 5, and 6 levels.  The latter cross sections for excitations from the n=3 levels occur just above the ionization threshold but are not very apparent in the total cross section.  For the DI component, a fit to the measurements has been obtained at energies below 650 eV.  For the EA component, only the cross sections from the n=2 levels are of importance and it has been necessary to rescale the magnitudes and shift the energy thresholds.  The energy threshold for the lowest energy EA component has been moved from 862 eV to 720 eV and the second component by a similar amount.  The magnitudes of the two EA components have been scaled by factors of 1.8 and 2.4.  The ratio of the present to the 2007 cross section is about 1.25 at an energy of 300 eV, just above the DI threshold.  At 500 eV the ratio is 1.06 and at energies above 1000 eV, the ratio is about 1.02.

\subsection{The argon isoelectronic sequence:  Fe IX} \label{sec:fe_9}

\begin{figure}[ht!]
\plotone{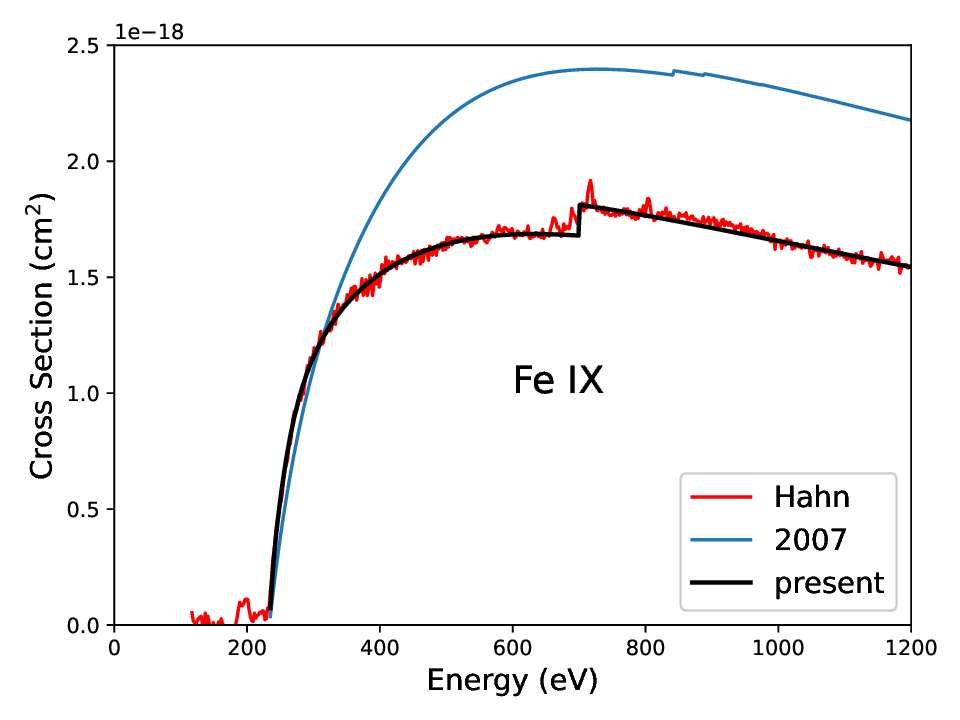}
\caption{Ionization cross sections of \ion{Fe}{ix}.  The label {\it Hahn} refers to \citet{hahn_fe_9}, {\it 2007} refers to \citet{ioniz}, and {\it present} to the results reported here.}
\label{fig:fe_9_cross}
\end{figure}

\ion{Fe}{ix} produces a very strong line at 171 \AA\ that is produced at about 8 $\times$ 10$^5$ K and shows characteristics of the solar upper transition region and low corona.  It comprises the primary component of narrow band EUV images such as the EIT on SOHO and the AIA on SDO.  A comparison of the measurements of \citet{hahn_fe_9}, \citet{ioniz} and the present results are shown in Fig. \ref{fig:fe_9_cross}.  The DI cross sections of \citet{ioniz} include ionization from the n=3 levels.  These have been replaced by a fit to the measurements at energies below about 700 eV and the result is that the DI cross section is reduced by about 40 per cent.  The EA components of the cross section in the original FAC calculations included excitations from the n=2 and n=3 levels and these cross sections were summed into three components with the lowest energy component at 842 eV.  \citet{hahn_fe_9} found the dominant EA component to have a threshold at about 650 eV and attributed it to excitations from the n=2 level to the 3d level.  This component was not included in the original FAC calculations.  This component is now modeled  by a fit to the measurements above 700 eV with a single EA component.  At an energy of 250 eV, just above the ionization potential of 233 eV, the ratio of the present cross section to the previous FAC cross section is about 1.5.  At 300 eV, the ratio is about 1.03, at 400 eV the ratio is about 1.04, and at 600 eV, near the peak of the DI component, the ratio is about 0.72.  In the region above 700 eV, the ratio is about 0.85.

\subsection{The potassium isoelectronic sequence:  Fe VIII} \label{sec:fe_8}

\begin{figure}[ht!]
\plotone{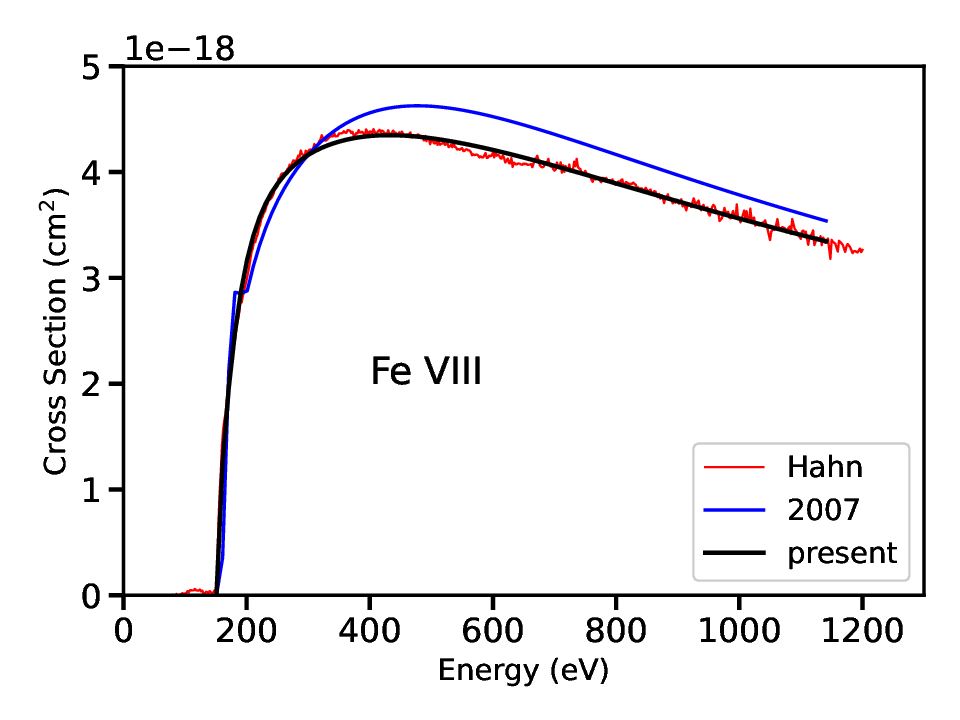}
\caption{Ionization cross sections of \ion{Fe}{viii}.  The label {\it Hahn} refers to \citet{hahn_fe_8}, {\it 2007} refers to \citet{ioniz}, and {\it present} to the results reported here.}
\label{fig:fe_8_cross}
\end{figure}

\ion{Fe}{viii} produces a number of lines at EUV wavelengths.  These are generally not particularly strong but do provide information on plasmas around 5 $\times$ 10$^5$ K.  Intensity ratios of several lines provide density diagnostics between 10$^6$ and 10$^{8}$ cm$^{-3}$.  A comparison of the measurements of \citet{hahn_fe_8}, \citet{ioniz} and the present results are shown in Fig. \ref{fig:fe_8_cross}.  Procedures were followed during the experimental measurements to ensure that as few ions in metastable states were present in the beam.  The authors state that ions in metastable states comprised only about 6\% of the beam.  The original FAC calculations included DI from the 3p and 3d shells and EA excitations from the 3p$^6$3d to the 3p$^5$3d nl levels with n=4,5,6.  The ionization potential is about 156 eV and the excitations to the n=4,5 and 6 autoionizing levels occur just above the ionization potential.  The various EA excitations were grouped into five EA cross sections with energies between 156 and 200 eV.  These are evident as step-like structures in the calculated ionization cross section just above threshold.  With the measurements of \citet{hahn_fe_8} now available, fits to the measurements were used to represent the cross sections.  These cross sections are shown in Fig. \ref{fig:fe_8_cross}.

\subsection{Ionization Rate Coefficients} \label{sec:rate}

The ionization rate coefficient, $R(T)$, is obtained by integrating the cross-section over a Maxwell-Boltzmann distribution, and is given by

\begin{equation}
R(T) = \int_{v_{IP}}^{\infty} v \, \sigma(E) \, f(v, T) \, dv
\label{equ:rate}
\end{equation}
where $v$ is the velocity of the electron, $E$ ($=\frac{1}{2} \, m \, v^2$) is the energy of the electron, and $f$ is the Maxwell velocity distribution at a temperature $T$.  The integration is carried out from the velocity of the electron with an energy equal to the ionization potential IP.

\begin{figure}[ht!]
\plotone{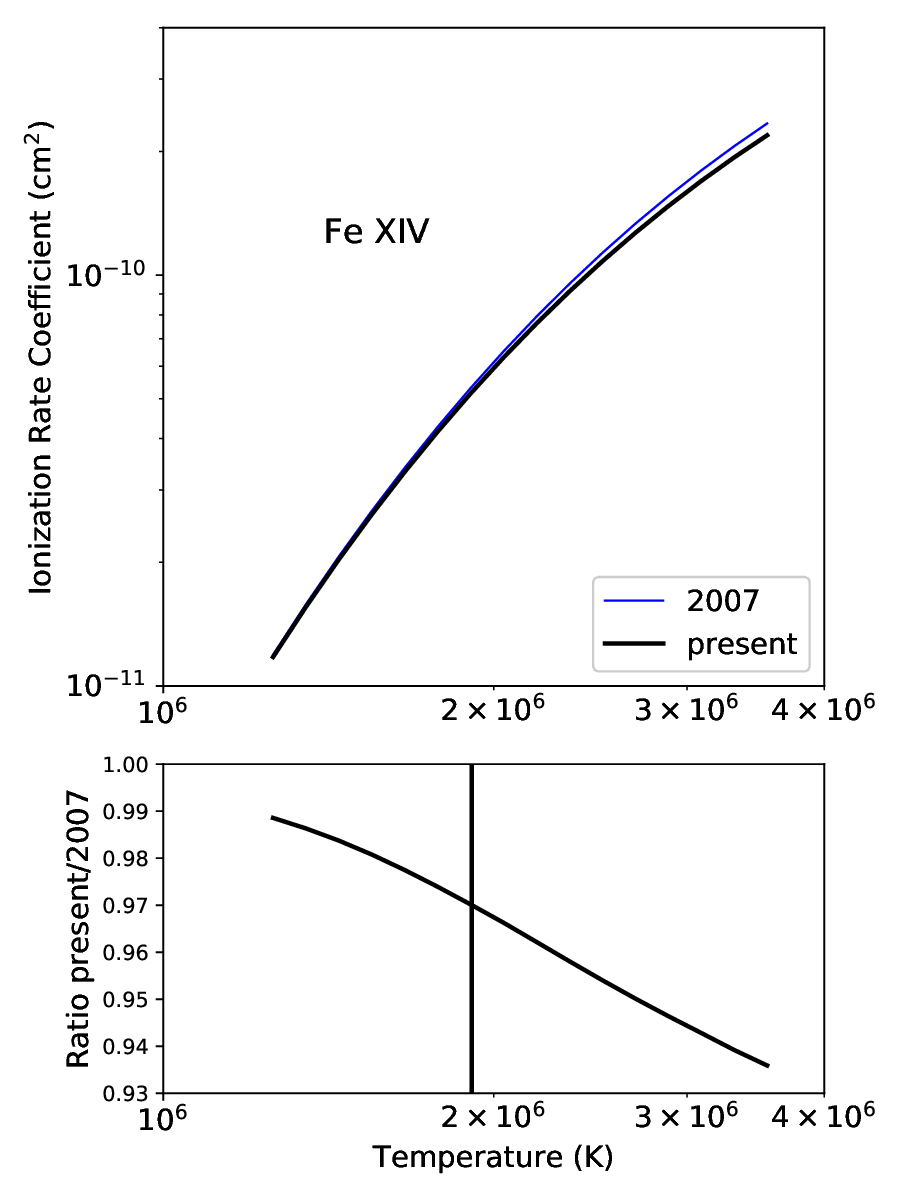}
\caption{Top:  Ionization rate coefficients of \ion{Fe}{xiv}.  The label {\it 2007} refers to \citet{ioniz}, and {\it 2022} to the present results.  Bottom:  The ratio of the present rate coefficients to those of {\it 2007}.}
\label{fig:fe_14_rate_ratio}
\end{figure}

\begin{figure}[ht!]
\plotone{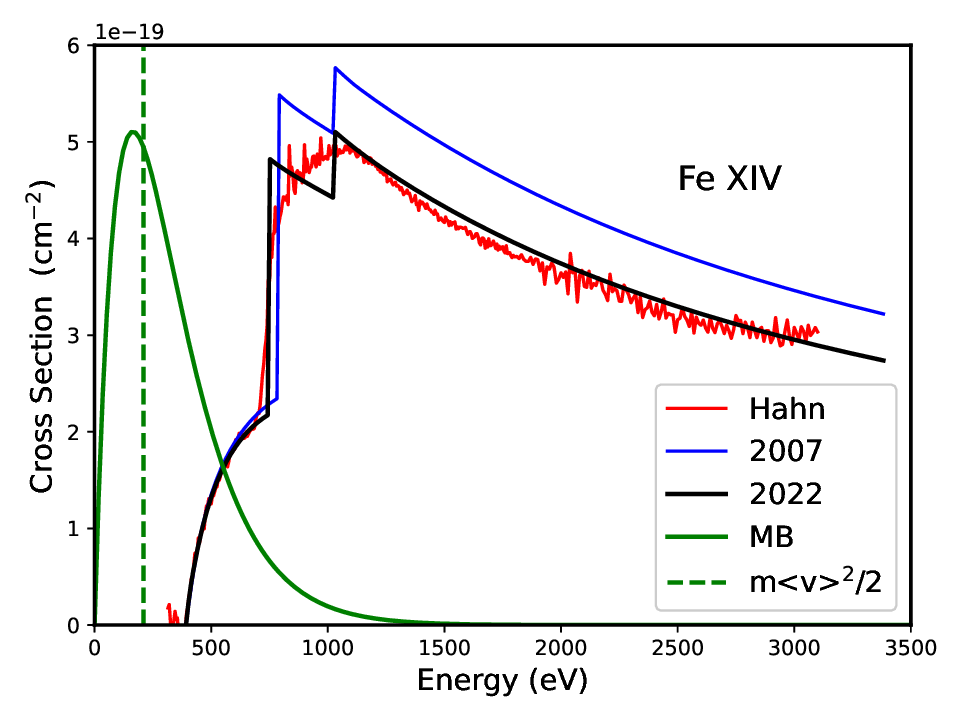}
\caption{Ionization cross sections of \ion{Fe}{xiv}.  The label {\it Hahn} refers to \citet{hahn_fe_14_17_18}, {\it 2007} refers to \citet{ioniz}, {\it present} to the results reported here.  Also plotted is the Maxwell-Boltzmann (MB) velocity distribution for a temperature of 1.91~MK and $\mathbf{m<v>^2/2}$ for the same temperature is displayed as a vertical line.}
\label{fig:fe_14_rate_ratio_mb}
\end{figure}

The \ion{Fe}{xiv} ionization rate coefficients for the present rates and those of \citet{ioniz} are shown in Fig. \ref{fig:fe_14_rate_ratio} together with the ratio of the rate coefficients.  The displayed temperature range approximately corresponds to where the \ion{Fe}{xiv} ionization fraction is within 0.01 of its peak value.  It can be seen that the revised rate coefficients do not show a significant difference from the previous values even though there are considerable differences in the ionization cross sections (see Fig. \ref{fig:fe_14_cross}).  The largest difference is in the EA components that lie well above the ionization potential.  In order to understand this lack of change, the Maxwell-Boltzmann distribution of electron velocities for a temperature $1.91 \times 10^6$~K has been plotted on Fig.~\ref{fig:fe_14_rate_ratio_mb}.  From this plot, it is clear that there is an insufficient number of electrons that are capable of participating in DI at energies where the cross section has changed since \citet{ioniz}, such as for \ion{Fe}{ix} nor in in EA processes where the cross section has changed.  The ratios of the new rate coefficients to those previously in CHIANTI for the ions in the present work are shown in Table~\ref{tab:rate_ratio}. In each case, the rate coefficients are calculated at the temperature of maximum ionization.

\begin{table}[ht!]
\begin{center}
\caption{The ratio of the ionization rate coefficients reported here to those of \citet{ioniz}}
\label{tab:rate_ratio}
\begin{tabular}{llc}
\hline
Ion & $T$ (10$^6$ K) & Ratio \\
\hline
\ion{S}{xiii}   &  2.7  &   1.11  \\
\ion{O}{v}      & 0.24  &   0.96 \\
\ion{Mg}{viii}  & 0.79  &   1.09 \\
\ion{Fe}{xviii} & 7.9   &   1.09 \\
\ion{Fe}{xvii}  & 5.6   &   1.06 \\
\ion{Fe}{xv}    & 2.2   &   0.99 \\
\ion{Fe}{xiv}   & 1.9   &   0.97 \\
\ion{Fe}{xiii}  & 1.7   &   1.09 \\
\ion{Fe}{xii}   & 1.6   &   1.17 \\
\ion{Fe}{xi}    & 1.3   &   1.09 \\
\ion{Fe}{x}     & 1.1   &   1.14 \\
\ion{Fe}{ix}    & 0.79  &   0.72 \\
\ion{Fe}{viii}  & 0.56  &   1.00 \\
\hline 
\end{tabular}
\end{center}
\end{table}

\subsection{Recombination rate coefficients for the phosphorus isoelectronic sequence} \label{sec:pseq}

Recently \citet{bleda_pseq} have reported radiative and dielectronic recombination coefficients for ions in the phosphorus sequence for ions with nuclear charge Z from 16 through 30 plus a few ions with higher nuclear charges.  These rates refer to the recombination of S-like ions to the P-like ions.  In the case of the recombination of \ion{Fe}{xii} to \ion{Fe}{xi} it was possible for the authors to compare their calculations with the measurements of \citet{novotny_dr}.  As the authors state, "Given the complexity of the problem, ..., the overall agreement, ..., is quite good."  \citet{bleda_pseq} provide fit parameters for their coefficients that are reproduced in the files released in the current CHIANTI release.

In general, the changes from the rates in the version 10.0 CHIANTI database are often not that large.  In the case of \ion{S}{ii}, the version 10.0 coefficients were taken from the work of \citet{mazzotta}  The agreement between the two is very good at high temperature but below $2 \times 10^4$~K, the rates of \citet{bleda_pseq} are significantly higher than those of \citet{mazzotta}.  \citet{mazzotta} often included the rates of \citet{shull} who provided fits to the the rates of \citet{jacobs_fe}.  As pointed out by \citet{savin_laming}, the calculations of Jacobs, and others, were computed with a code that only considered LS levels.  At low temperatures, the strongest process for dielectronic recombination proceeds through a dielectronic excitation involving no change in the principle quantum number ($\Delta$N = 0) in the core electrons.  This is followed by a stabilizing ($\Delta$N = 0) radiative transition.  These transitions can not be included in LS models and consequently the dielectronic recombination rate at low temperatures is underestimated.

However, the differences are mostly of importance for photoionized plasmas.  For high temperature dielectronic recombination ($\Delta$N $>$ 0), of \ion{Ca}{vi} the new DR rates of \citet{bleda_pseq} are about 50\% lower than the rates of \citet{shull} that were included in CHIANTI version 10.0.  For \ion{Ni}{xiv} \citet{bleda_pseq} show considerable difference between their DR rates and those of \citet{mazzotta}.  Version 10.0 of CHIANTI still included many of the DR rates of \citet{shull} for the phosphorus isoelectronic sequence and these are often roughly comparable to the new rates of \citet{bleda_pseq} at higher temperatures.

\subsection{A revised ionization balance}\label{sec:ioneq}

\begin{figure}[ht!]
\plotone{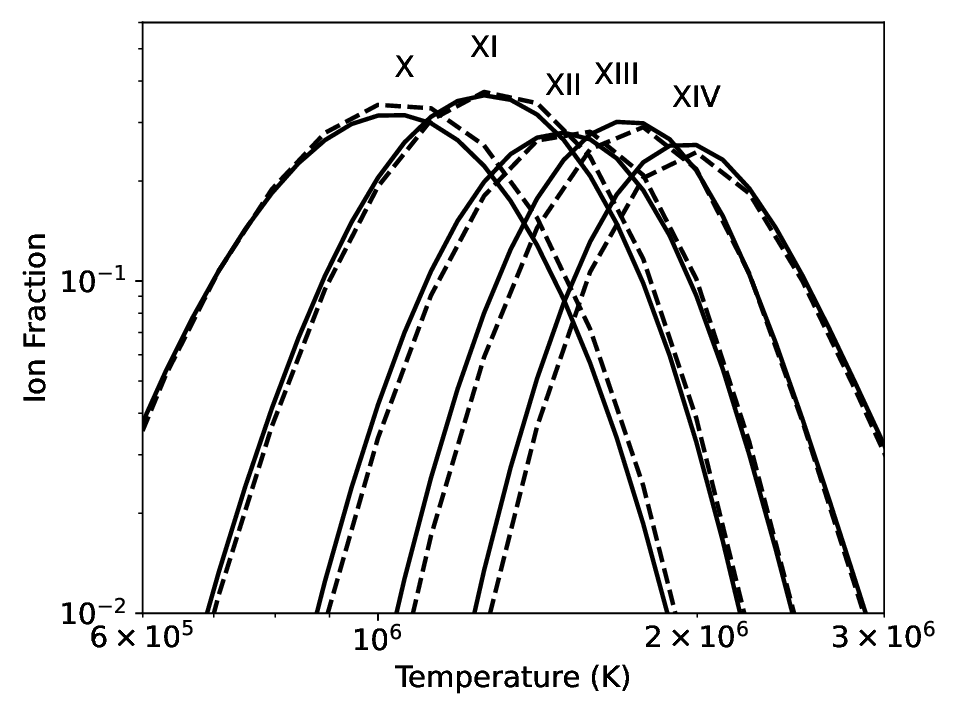}
\caption{The revised ionization balance for Fe ions for stages X through XIV.  The full line represents the present calculation and the dashed line the version 10.0 calculation}
\label{fig:fe_ioneq}
\end{figure}

With the revised set of ionization and recombination rate coefficients, an updated ionization balance has been calculated and will be distributed as the default ionization equilibrium for CHIANTI 10.1. Figure~\ref{fig:fe_ioneq} compares the new ionization fraction curves for the iron ions \ion{Fe}{x--xiv} with those from CHIANTI 10. Comparing all of the ionization fraction curves between the new and previous tabulations, there are 11 ions with $\log\,T_\mathrm{max}$ values shifted by 0.05~dex and seven ions with peak ionization fractions that are $>10$\%\ different compared to the previous ionization fraction curves. The latter include \ion{V}{viii,ix}, \ion{Sc}{vi,vii}, \ion{Ca}{vi} and \ion{K}{v}, where the differences are due to the updated DR rates. The largest difference is a 30\%\ increase for \ion{S}{i}.

\section{New and updated atomic models}\label{sec:models}

\subsection{Hydrogen isoelectronic sequence: C VI}

An error in the proton rate coefficients used for \ion{C}{vi} was identified by C.~M.~Gunasekera and has been corrected. The rate coefficients for the transitions between the three levels of the $n=2$ configuration were computed by \citet{1987PhRvA..35.4085Z} and added to CHIANTI Version 6 \citep{paper9}. \citet{1987PhRvA..35.4085Z} tabulated their rate coefficients with a scaling of the atomic number, $Z$, to the power of three. In removing the scaling, the numbers were inadvertently multiplied by  $Z^3$ instead of divided, leading to rate coefficients too large by a factor $4.7\times 10^4$.

The proton rates for \ion{C}{vi} provide a depopulation mechanism for the $2s$ $^2S_{1/2}$ metastable level in addition to the two-photon decay mechanism. The erroneous rates resulted in proton excitation being the dominant depopulation mechanism. With the correct rates, the two-photon process becomes the dominant process. As a consequence, the two-photon continuum for \ion{C}{vi} is now more than a factor 1000 stronger than previously for a temperature of 1~MK and an electron number density of $10^{10}$~cm$^{-3}$.

A second effect is that the erroneous proton rates resulted in the populations of the $2p$ $^2P_{1/2,3/2}$ levels being enhanced by 44\%\ and 10\%\, respectively, for a temperature of 1~MK and an electron number density of $10^{10}$~cm$^{-3}$. The Ly-$\alpha$ line for \ion{C}{vi}, which is a self-blend of the transitions from the $^2P$ levels to $1s$ $^2S_{1/2}$, is consequently 18\%\ weaker in the new \ion{C}{vi} model compared to CHIANTI 10.

\subsection{Boron isoelectronic sequence: O IV}

Errors were found in the energy levels for this ion such that experimental energies had been inadvertently assigned to the wrong levels. In particular, the energies for levels 27, 28 and 51  had been assigned to levels 29, 30 and 52, respectively, and vice versa. The energies have been updated and a new set of wavelengths have been derived from the updated energies. No other changes were made to the \ion{O}{iv} model.

\subsection{Carbon isoelectronic sequence}


\subsubsection{Ca XV}

Updated atomic models for the entire carbon isoelectronic sequence were added to CHIANTI 10. However, it was found that the \ion{Ca}{xv} effective collision strengths were omitted from the update. This has been rectified for  CHIANTI 10.1.

\subsection{Nitrogen isoelectronic sequence}

\citet{2020A&A...643A..95M}  presented a large-scale scattering calculation using the
$R$-matrix intermediate coupling frame transformation (ICFT) method,
to calculate effective collision strengths of N-like ions from \ion{O}{ii}
to \ion{Zn}{xxiv}. 725 fine-structure levels were included
in both the configuration interaction target and close-coupling collision expansion, for states up to $n=5$.
We are providing new models based on these calculations for the
\ion{O}{ii}, \ion{Si}{viii}, \ion{Ar}{xii} and \ion{Ca}{xiv}. As shown in \citet{2020A&A...643A..95M}, the new rates for these ions are 
significantly different than those previously available in CHIANTI, and they affect some line ratio diagnostics.

For the experimental energies we have used a combination of
NIST \citep{NIST_ASD} values and those that were assessed in the previous CHIANTI
versions.  Only bound states have been retained for O II and Si VIII, as these ions
do not produce strong satellite lines.

For  \ion{O}{ii}, we have replaced the effective  collision strengths 
for the first five states of the ground configuration with those 
from \citet{2007ApJS..171..331T}, up to 10$^5$ K, as we consider them to be more
accurate at low temperatures. We note that the values of the
two calculations at 10$^5$ K differ by only a few percent,
as shown in \citet{2020A&A...643A..95M}.

\citet{2020A&A...643A..95M} used {\sc autostructure} (AS, \citealt{badnell:2011}) to define the target for the scattering calculations.  Generally, the transition probabilities of the lower states calculated with AS are not as accurate as those obtained with larger-scale calculations, such as with the Multiconfiguration Dirac-Hartree-Fock (MCDHF) GRASP2K codes, described in \cite{jonsson_etal:2013}.

For  \ion{O}{ii}, \ion{Si}{viii}, and \ion{Ar}{xii}, we have replaced the AS $A$-values of the transitions within the 
2s$^2$ 2p$^3$, 2s 2p$^4$ configurations with the 
Breit-Pauli calculations of \cite{tff:2002}. 
For \ion{Ca}{xiv},  we have replaced the $A$-values of the transitions within the 
2s$^2$ 2p$^3$, 2s 2p$^4$ configurations with the 
many-body perturbation theory (MBPT) calculations
by \citet{2016ApJS..223....3W}.  The AS radiative data only differed by typically 10--20\%, compared to the MCHF or MBPT values.

\subsection{Oxygen isoelectronic sequence}

\cite{mao_etal:2021A&A...653A..81M}
 performed a large-scale scattering calculation using the
$R$-matrix ICFT method,
to calculate effective collision strengths of O-like ions from
\ion{Ne}{III} to \ion{Zn}{XXIII} over a wide range of temperatures.
The targets included 630 fine-structure levels up to $nl=5d$ for each ion.
As discussed in \citet{mao_etal:2021A&A...653A..81M}, significant differences with the
data that are present in CHIANTI were found for a few ions
important for astrophysical applications. They include \ion{Si}{vii}, \ion{S}{ix}, \ion{Ar}{xi}, and \ion{Ca}{xiii},
for which CHIANTI had limited models and effective collision strengths
calculated with the distorted wave approximation.
Lines in the EUV, UV and near infrared have significantly
different predicted intensities with the new atomic data.
For this reason, we have included new models for these ions.
Work is in progress to update the models for the other ions,
which will be made available in a future CHIANTI version.

For the experimental energies we have used a combination of
NIST values and those that were assessed in the previous CHIANTI
versions.  Only bound states have been retained, as the model of the
autoionizing states requires significant additions. 

For \ion{Si}{vii} and \ion{S}{ix} we have replaced the new AS A-values of the transitions within the $n=2$ complex (ground configuration 2s$^2$ 2p$^4$, 2s 2p$^5$ and 2p$^6$, ten lowest states)
with those calculated with the Breit-Pauli codes by \cite{tff:2002}, noting that the \citet{mao_etal:2021A&A...653A..81M} A-values differed by typically 10--20\% only.  For \ion{Ar}{xi} and \ion{Ca}{xiii}  we have replaced the AS values with those calculated by \cite{song_etal:2021ADNDT.13801377S} with the  GRASP2K codes.  Differences with the AS values are about 10\%.

\subsection{Silicon isoelectronic sequence: Fe XIII}

\cite{zhang_etal:2021ApJS..257...56Z} carried out a large-scale MCDHF atomic
structure calculation for  \ion{Fe}{xiii}, providing theoretical energies that are
generally  close to the experimental energies, typically within a few hundreds of cm$^{-1}$.  
\cite{zhang_etal:2021ApJS..257...56Z}  reviewed previous studies and confirm all 
the new identifications suggested by \cite{delzanna:2012_sxr1}
(based on the \citet{delzanna:2011_fe_13} calculations) for the 
$n=4 \to n=3$ transitions, with one exception.
This involves the strongest line 
of this transition array, with the new calculations. 
\cite{delzanna:2012_sxr1} pointed out three possible 
wavelengths, at 76.113, 76.507, and
76.867~\AA, and chose the 76.507~\AA. However, this wavelength differs by 0.35~\AA\ from the ab-initio MCDHF value, while the 76.867~\AA\ line only by 0.015~\AA,
so \cite{zhang_etal:2021ApJS..257...56Z} suggested this new 
identification, which we adopt. 
Several new tentative identifications were also suggested by 
\cite{zhang_etal:2021ApJS..257...56Z} but are not adopted here, as further experimental studies are needed to confirm them.

A few weak EUV transitions from the 3s 3p$^2$ 3d levels were tentatively identified 
by \cite{delzanna:2011_fe_13} and assigned experimental wavelengths, for their 
potential importance in analyzing Hinode EIS observations. However, the analysis 
carried out by \citet{zhang_etal:2021ApJS..257...56Z} provides evidence for alternative 
identifications. Some of them have been adopted for the present version, while
other tentative identifications have not been included. 
However, we are providing as `best' energies those from Zhang et al.,
hence the expected wavelengths should be close to the experimental ones.

\subsection{Argon isoelectronic sequence: Fe IX}

\citet{2022ApJ...936...60R} provided a new set of experimental energies for the $3p^43d^2$ and $3p^54f$ configurations and these have been added to CHIANTI with the modifications described below.

The CHIANTI 10 model had experimental energies for all 12 of the $3p^54f$ fine-structure levels from \citet{2012A&A...537A..22O}, and these have been updated with the values from \citet{2022ApJ...936...60R}. Agreement is good for nine of the levels, but larger differences of between 160 and 560~cm$^{-1}$ are found for the remaining levels due to differences in line identifications---see \citet{2022ApJ...936...60R} for more details.

Comparisons of the $A$-values listed by \citet{2022ApJ...936...60R} with those in the CHIANTI model \citep[from][]{2014A&A...565A..77D} showed that it was necessary to swap the \citet{2022ApJ...936...60R} labels for the $^1F_3$ and $^3F_3$ levels, and the $^1G_4$ and $^3F_4$ levels to get agreement.
Both sets of authors found that these levels are strongly mixed, hence the $LSJ$ labels are not accurate descriptors for them. The \citet{2014A&A...565A..77D}  and \citet{2022ApJ...936...60R} $A$-values are in very good agreement after this modification.

The $3p^43d^2$ configuration has 111 fine-structure levels, but only five previously had experimental energies. Sixteen new experimental energies have been taken from \citet{2022ApJ...936...60R}, and the five existing energies have been updated, with only small changes.  \citet{2022ApJ...936...60R} provided experimental energies for nine additional levels, but three of these were not used as they were listed as `questionable.' Six further energies were not used either because the identifications were not consistent with predictions from the CHIANTI atomic model, or because the identifications were uncertain due to line blending. Full details of how these decisions were made are given in \citet{peter_young_2023_7803672}.

\citet{2009ApJ...707..173Y} performed a study of \ion{Fe}{ix} lines observed by the Extreme Ultraviolet Imaging Spectrometer \citep[EIS][]{2007SoPh..243...19C}, and they identified seven lines that could be attributed to \ion{Fe}{ix} but for which it was not possible to assign atomic transitions. Two of these lines were independently identified as \ion{Fe}{ix} lines by \citet{2009A&A...508..501D}, but also without atomic transitions. \citet{2014A&A...565A..77D} provided transition information for two lines, and \citet{2022ApJ...936...60R} identified six of them. The two works agree for the line at 192.63~\AA, but differ for the line at 194.80~\AA. The \citet{2022ApJ...936...60R} identification is supported by multiple lines from the same upper level that are measured in laboratory spectra.

The CHIANTI wavelengths in the present version have been derived from the new energy levels.

\subsection{Calcium isoelectronic sequence: Fe VII}

Experimental energies have been updated using the compilation of \citet{2022ApJS..258...37K} that provided energies for previously unassigned levels and updated energies of known levels. A number of levels are highly mixed, particularly for the $3p^53d^3$ configuration and it was not always possible to match the \citet{2022ApJS..258...37K} level with a level in the existing CHIANTI model. In these cases the \citet{2022ApJS..258...37K} level was not used. Details on which level energies were not chosen for the CHIANTI model and how the decisions were made are described in \citet{peter_young_2023_7799540}. In summary, the new model has 17 levels with newly-assigned experimental energies, and 10 levels for which the updated energy is at least 100~cm$^{-1}$ different from the previous energy. The wavelengths in the present version have been updated with the new energies.  However, as pointed out in the previous CHIANTI release paper,
some inconsistencies remain between the predicted intensities of the strongest 
lines and those observed by Hinode EIS.  \cite{delzanna:09_fe_7} used an alternative set of atomic data and identifications to find relatively good agreement with observations. As different calculations provide very different 
energies and rates (even branching ratios), the solution of this complex 
problem will require further calculations and further assessments.

\section{Elemental Abundances}

The CHIANTI software uses solar photospheric elemental abundances as the default when computing synthetic spectra. For CHIANTI versions 9 and 10, the photospheric abundances were stored in the file \textsf{sun\_photospheric\_2015\_scott.abund}. This contains the abundances of \citet{2009ARA&A..47..481A}, supplemented with  values from \citet{2015A&A...573A..26S, 2015A&A...573A..25S} and \citet{2015A&A...573A..27G} for some elements.

For CHIANTI 10.1 the default abundance file---named \textsf{sun\_photospheric\_2021\_asplund.abund}---contains data from the  compilation of \citet{2021A&A...653A.141A}.  These authors provide a comprehensive and detailed review of the photospheric abundance of all elements up to uranium, complementing recent re-evaluations and new measurements in the literature for the abundances of many elements with original results. In both cases, improved values were obtained mostly from the coupling of photospheric observed spectra with 3D, NLTE hydrodynamical simulations of the outer convective zone and atmospheric layers of the Sun, and improved atomic data and collisional rates for many species. Additional data were taken from laboratory analysis of meteorites, return samples from the Genesis mission, helioseismological data, and sunspot observations.

In comparison to the previous default abundance set, the \citet{2021A&A...653A.141A} abundances show differences of 10\%\ or more for four elements: lithium ($-19$\%), neon ($+35$\%), chlorine ($-19$\%) and titanium ($+10$\%). The most significant for the spectral modeling of ionized plasmas is the neon abundance.

The new dataset provides a Ne/O relative abundance ratio of 0.23, larger than the previous value of 0.17. The new value is in line with recent spectroscopic re-evaluations of the Ne/O abundance ratio in the transition region \citep{2018ApJ...855...15Y} and in the solar corona during solar minimum \citep{2015ApJ...800..110L}. The Ne/O ratio, and the absolute abundances of both Ne and O are very important for solving the current discrepancy between helioseismology determinations of the structure of the solar interior with model predictions, that need accurate absolute abundance measurements to calculate solar interior opacities \citep{2021LRSP...18....2C}.

We finally note that the coronal abundance data sets provided in earlier CHIANTI versions (e.g.,
\textsf{sun\_coronal\_1992\_feldman.abund} and \textsf{sun\_coronal\_2012\_schmelz.abund}) were largely based on solar active region observations 
 where the relative abundances of low vs.\ high first ionization potential (FIP) 
 elements showed an increase (the so-called FIP bias), compared to their
 (relative) photospheric values 
 \citep[see, e.g. the reviews by ][]{2015LRSP...12....2L,delzanna_mason:2018}.
The FIP bias was found to have averaged values around 3--4. 
Those coronal abundance data sets were obtained by 
 taking into account those averaged values and applying empirical corrections (the FIP bias) to the older photospheric abundance data sets. 
 The previous coronal abundance datasets are therefore not consistent 
 with the new photospheric abundance file discussed here. We have therefore introduced a new coronal abundance file \textsf{sun\_coronal\_2021\_chianti.abund} that is derived from the new photospheric abundance file by multiplying abundances for low-FIP elements (FIP $\le$ 10~eV) by a factor $10^{0.5}$. This file is provided as a representative coronal abundance file that may be useful for investigating how abundances affect synthetic spectra, for example. Users should be aware that the magnitude of the FIP bias has been found to vary amongst different solar structures, and also with temperature.

 Abundance files (both photospheric and coronal) previously distributed with CHIANTI will continue to be available but are moved to the \textsf{abundance/archive} directory within the database.

\section{Conclusions}  \label{sec:conclusions}

This article has described the most recent updates to the CHIANTI atomic database that will be distributed as version 10.1. Ionization rates for 13 ions have been updated, as well as the complete set of dielectronic recombination data for the phosphorus isoelectronic sequence. The core datasets of energy levels, radiative decay rates and electron excitation rates have been updated for 15 ions.

\citet{ioniz} assembled  the experimental measurements of ionization cross sections available at that time. These were supplemented with calculations of the ionization cross sections with the FAC package \citep{fac} where measurements did not exist or were unreliable.  Together with recombination rates from the scientific literature, version 9.0 of the CHIANTI atomic database was assembled with a full set of rates covering all ions of all elements from H through Zn.  Since that time, there have been a number of new experimental measurements of ionization cross sections, as well as new calculations of recombination rates.  The new ionization measurements have been used to revised the CHIANTI parameterization of these cross sections and these have been inserted into the latest version of the CHIANTI atomic database (Version 10.1).  In most cases, the differences between the revised rates and the previous CHIANTI rates is not that great.  The parameters for recombination of \citet{bleda_pseq} have also been inserted into the latest version of CHIANTI.  In addition, a new ionization equilibrium has been calculated from these revised rates.  

A number of atomic models have been updated.  For the nitrogen isoelectronic sequence, new models for the \ion{O}{ii}, \ion{Si}{viii}, \ion{Ar}{xii} and \ion{Ca}{xiv} ions are now based on the recent calculations of \citet{2020A&A...643A..95M}.  For the oxygen isoelectronic sequence,  the calculations of \cite{mao_etal:2021A&A...653A..81M} have allowed us to improve the previously limited models for \ion{Si}{vii}, \ion{S}{ix}, \ion{Ar}{xi}, and \ion{Ca}{xiii}.
 
 All of these are now included in the latest release of the CHIANTI atomic database version 10.1.

\begin{acknowledgments}

We thank Dr. Stefan Schippers for supplying the measurements of \citet{bleda_pseq} in a machine-readable format.  KPD and PRY work has been supported by NASA grants 80NSSC21K0110 and 80NSSC21K1785. 
GDZ acknowledges support from STFC (UK) via the consolidated grants 
to the atomic astrophysics group (AAG) at DAMTP, University of Cambridge (ST/P000665/1. and ST/T000481/1).
EL has been supported by NASA grants 80NSSC22K0750 and 80NSSC20K0185.

\end{acknowledgments}

\bibliography{ioniz}{}
\bibliographystyle{aasjournal}

\end{document}